\documentclass[a4paper,conference]{IEEEtran}

\ifCLASSINFOpdf

\else

\fi

\usepackage{amsmath,amssymb}
\usepackage{graphicx}
\usepackage{xspace}
\usepackage{color}
\usepackage{epstopdf}
\usepackage{algorithm}
\usepackage{algorithmic}
\usepackage{setspace}
\usepackage[font=small,labelfont=bf]{caption}
\usepackage{subcaption}
\usepackage{pgfplots, tikzscale}
\usetikzlibrary{plotmarks}
\usetikzlibrary{plotmarks,patterns}
\pgfplotsset{compat=newest}
\usepackage{multirow}
\usepackage{makecell}
\definecolor{darkpastelgreen}{rgb}{0.01, 0.75, 0.24}
\definecolor{azure}{rgb}{0.0, 0.5, 1.0}
\usepackage{booktabs} 
\usepackage[hyphens]{url}
\graphicspath{ {figures/} }
\newcommand{\systemname}{DFT-s-NL\xspace}
\def\thesubsubsectiondis{\unskip\arabic{subsubsection})}

 \usetikzlibrary{external}
 \usepackage{tikz}

\newcommand\copyrighttext{%
	\footnotesize 2026 IEEE. Personal use of this material is permitted. Permission from IEEE must be obtained for all other uses, in any current or future media, including reprinting/republishing this material for advertising or promotional purposes, creating new collective works, for resale or redistribution to servers or lists, or reuse of any copyrighted component of this work in other works.}
\newcommand\copyrightnotice{%
	\begin{tikzpicture}[remember picture,overlay]
		\node[anchor=south,yshift=10pt] at (current page.south) {\fbox{\parbox{\dimexpr\textwidth-\fboxsep-\fboxrule\relax}{\copyrighttext}}};
	\end{tikzpicture}%
}

\begin{document}
\bstctlcite{IEEEexample:BSTcontrol}





\title{\fontsize{18.9}{18.9}\selectfont Aggressive  Non-Orthogonal Transmission with DFT-s-OFDM for Direct Device-to-Satellite Communications}

	
\author{
	\IEEEauthorblockN{Chathura Jayawardena\IEEEauthorrefmark{1} and Konstantinos Nikitopoulos\IEEEauthorrefmark{1}\IEEEauthorrefmark{2}}
	\IEEEauthorblockA{\IEEEauthorrefmark{1}Wireless Systems Lab, 6GIC, University of Surrey, Guildford GU2 7XH, UK}
	\IEEEauthorblockA{\IEEEauthorrefmark{2}Noema Signal Labs, Cambridge, UK}
	\vspace{-24pt}
}	
	
\maketitle
\copyrightnotice
\begin{abstract}

Direct Device-to-Satellite (D2S) communications promise global connectivity to unmodified user equipment (UE), extending coverage beyond terrestrial networks. Realizing this promise is fundamentally challenging: severe path loss and limited UE transmit power push uplink SNRs far below terrestrial norms, while suitable spectrum remains scarce. Together, these constraints impose a spectral-efficiency (SE) bottleneck, and under such conditions the efficiency of the UE power amplifier becomes critical, jointly governing transmit power and battery life. To improve UE-side power efficiency, 3GPP has adopted Discrete Fourier Transform-spread OFDM (DFT-s-OFDM) as an optional uplink waveform, exploiting its substantially lower Peak-to-Average Power Ratio (PAPR) relative to OFDM. To break the SE bottleneck, we show that \emph{aggressive} non-orthogonal transmission, in which the number of concurrent users exceeds the number of receive antennas by more than $2\times$, can unlock substantial capacity gains that remain entirely unexploited. Realising these gains, however, requires receiver architectures that, to the best of our knowledge, have not yet been developed. DFT-s-OFDM intensifies the difficulty: the DFT spreading couples signal components across subcarriers, inflating the effective dimensionality of the detection problem. We address both challenges with a novel receiver design that jointly exploits the SE gains of aggressive non-orthogonal transmission and the power-efficiency benefits of DFT-s-OFDM. Simulations under realistic channel-estimation errors and high-mobility Doppler show that the proposed scheme achieves $2\times$ the SE of baseline, surpasses recent nonlinear MIMO receivers by $40\%$ at $15\%$ of their complexity, and reduces PAPR by up to 6 dB relative to DFT-s-OFDM MIMO and 11 dB relative to OFDM.

\end{abstract}

\vspace{-4pt}
\section{Introduction}
\vspace{-2pt}
Direct Device-to-Satellite (D2S) communication, recently standardized within the 3GPP Non-Terrestrial Network (NTN) framework, enables unmodified User Equipment (UE) to connect directly to orbiting satellites, promising truly ubiquitous coverage and marking a paradigm shift in wireless access. Realizing this vision, however, poses fundamental challenges. The combination of extreme propagation losses and the constrained transmit power of handheld devices depresses uplink Signal-to-Noise Ratios (SNRs) to levels well below those of terrestrial networks, while the spectrum with the propagation characteristics required to sustain the link is exceedingly scarce. Together, these constraints give rise to a severe spectral-efficiency bottleneck. 

To enhance UE-side power efficiency, which jointly governs achievable transmit power and battery life, 3GPP has adopted Discrete Fourier Transform-spread OFDM (DFT-s-OFDM) as an optional uplink waveform, motivated by its substantially lower Peak-to-Average Power Ratio (PAPR) compared with conventional OFDM. By inserting a DFT-precoding stage ahead of OFDM modulation, DFT-s-OFDM produces a single-carrier-like signal whose PAPR is markedly reduced, as summarised in Table~\ref{tab:PAPR}. The reduced PAPR allows the power amplifier to operate with smaller back-off, closer to its saturation point, directly improving both the link budget and the energy efficiency of the UE.

Despite these power-efficiency advantages, DFT-s-OFDM with orthogonal multi-user access exhibits limited spectral efficiency and connectivity, posing a bottleneck in both non-terrestrial and terrestrial deployments. Non-orthogonal transmission schemes, notably Multi-User MIMO (MU-MIMO) and Non-Orthogonal Multiple Access (NOMA), offer a natural means of addressing this limitation and promise substantial gains in capacity and connectivity.  Realizing these gains in DFT-s-OFDM, however, is considerably more challenging than in conventional OFDM. The DFT-precoding stage spreads each data symbol across the entire allocated bandwidth, coupling data symbols across subcarriers and breaking the per-subcarrier independence that underpins most standard MIMO receivers. As a result, near-optimal receivers capable of realizing the theoretical spectral efficiency of non-orthogonal DFT-s-OFDM transmissions remain largely unexplored.

	\begin{table}[t]
	\fontsize{7}{8}\selectfont
	\caption{PAPR comparison of DFT-s-OFDM and OFDM waveforms at the
		$10^{-3}$ CCDF level. Values are representative of typical 5G NR
		uplink configurations
.}
	\centering
	\begin{tabular}{lcc}
		\toprule
		\textbf{Modulation} & \textbf{DFT-s-OFDM} & \textbf{OFDM} \\
		\midrule
		$\pi/2$-BPSK   & $\sim$1--2 dB     & $\sim$10.5--12 dB \\
		QPSK (4-QAM)   & $\sim$5.5--6 dB   & $\sim$10.5--12 dB \\
		16-QAM         & $\sim$6.5--7 dB   & $\sim$10.5--12 dB \\
		64-QAM         & $\sim$7.5--8 dB   & $\sim$10.5--12 dB \\
		256-QAM        & $\sim$8--8.5 dB   & $\sim$10.5--12 dB \\
		\bottomrule
	\end{tabular}
	\label{tab:PAPR}
	\vspace{-18.5pt}
\end{table}

To enable receiver processing in DFT-s-OFDM-based MIMO systems, the work in  \cite{SC-FDMblccirc} proposes a joint subcarrier processing scheme. A key limitation of this approach, however, is the need for high-dimensional matrix inversions, resulting in complexity that scales cubically with the number of jointly processed subcarriers, even when only linear processing is considered. To address this, the work in \cite{DL_GAMP} adapts the message passing algorithm to DFT-s-OFDM-based MIMO systems by introducing DFT and IDFT operations at each iteration. Nevertheless, this approach has two notable drawbacks.  First, message passing algorithms rely on sparse graphical models to achieve near-optimal performance, a condition that is not satisfied by general MIMO channel matrices, and typically requires specifically designed sparse signals as in Code Domain NOMA schemes \cite{SCMA,SCMA2}. Second, the algorithm typically requires a large number of iterations (e.g., 50), making the cumulative overhead of the per-iteration DFT and IDFT operations a significant contributor to overall complexity.

In this work, we propose a framework that reframes this spectral and power-efficiency challenge by jointly exploiting two complementary properties: the capacity gains achievable through aggressive non-orthogonal transmissions and the low PAPR of DFT-s-OFDM. Our contributions are threefold. \textit{First}, we show that the low-SNR regime characteristic of D2S and coverage-limited terrestrial deployments (e.g., cell-edge UEs) admits a previously unexplored capacity-growth opportunity, one that ``aggressive'' non-orthogonal transmissions can uniquely exploit. 
In particular, exceeding the conventional NOMA limit of two UEs per frequency element and receive antenna yields substantial additional capacity, independent of spatial multiplexing gains. Power-efficiency benefits accrue on three complementary fronts, at both the UE and the base station: the inherently low PAPR of DFT-s-OFDM at the UE, further PAPR reduction afforded by low-order or near-constant-envelope modulation, and fewer receive antennas and RF chains required at the base station. Realizing these gains, however, demands receiver architectures capable of handling such aggressive overloading, and to the best of our knowledge, no existing design meets this requirement. The challenge is further compounded by DFT-s-OFDM, whose spreading operation increases the dimensionality of receiver processing. \textit{Second}, we formulate the joint detection problem that translates these theoretical gains into realized throughput, and identify the computational barriers that render its direct implementation intractable. \textit{Third}, we develop a reduced-complexity receiver that circumvents these barriers while closely approximating the optimal joint detector, making the spectral and power-efficiency gains of aggressive non-orthogonal DFT-s-OFDM practically attainable. Notably, the proposed scheme achieves up to $2\times$ the spectral efficiency of the baseline, even under practical channel-estimation errors and the high-mobility Doppler impairments characteristic of LEO-based D2S links. It also delivers $40\%$  higher spectral efficiency at only $15\%$  of the complexity of recent nonlinear MIMO receivers \cite{DL_GAMP}, and reduces UE-side PAPR by up to 6 dB relative to DFT-s-OFDM-based MIMO transmissions and up to 11 dB relative to OFDM-based systems.

	\vspace{-9pt}
\section{DFT-s-OFDM Based Non-Orthogonal Signal Transmissions}
\label{sec:system_model}
\vspace{-8pt}
We consider the uplink in which $K$ single-antenna UEs transmit concurrently to a base station equipped with $M$ receive antennas over the same $N$ allocated subcarriers. Each UE employs DFT-s-OFDM with a cyclic prefix (CP) longer than the maximum channel delay spread, so that each OFDM symbol experiences a circulant time-domain channel.

Let $s_{k,n} \in \mathcal{O}$ denote the data symbol of UE $k$ at the $n^{th}$ DFT input, where $\mathcal{O}$ is the transmit symbol constellation and $\mathbb{E}\{|s_{k,n}|^2\}=P_t$ with $\mathbb{E}\{\}$ denoting the expected value. To expose the per-subcarrier structure of the transmission, we stack the symbols across users at each DFT index as $\mathbf{s}_n = [s_{1,n},\, s_{2,n},\, \ldots,\, s_{K,n}]^T \in \mathbb{C}^{K \times 1}$, and concatenate them into
\begin{equation}
	\bar{\mathbf{s}} = \begin{bmatrix}\mathbf{s}_1^T & \mathbf{s}_2^T & \cdots & \mathbf{s}_N^T\end{bmatrix}^T \in \mathbb{C}^{NK \times 1}.
\end{equation}
An $N$-point DFT is then applied independently to each UE's symbol sequence, yielding the frequency-domain transmit vector
\begin{equation}
	\bar{\mathbf{x}} = (\mathbf{F}_N \otimes \mathbf{I}_K)\,\bar{\mathbf{s}} \in \mathbb{C}^{NK \times 1},
	\label{eq:dft_precoding}
\end{equation}
where $\mathbf{F}_N \in \mathbb{C}^{N \times N}$ is the unitary DFT matrix and $\otimes$ denotes the Kronecker product. Partitioning the result as $\bar{\mathbf{x}} = [\mathbf{x}_1^T,\, \mathbf{x}_2^T,\, \ldots,\, \mathbf{x}_N^T]^T$ with $\mathbf{x}_n \in \mathbb{C}^{K \times 1}$, the $k$-th entry of $\mathbf{x}_n$ is the DFT precoded symbol that UE $k$ places on the $n$-th subcarrier. Each UE subsequently maps its $N$ DFT precoded symbols onto the assigned subcarriers within a larger IFFT grid and appends a CP before transmission. Since the IFFT/FFT pair together with the CP diagonalize the circulant time-domain channel, they are omitted from the baseband representation without loss of generality.
After CP removal and OFDM demodulation at the base station, the received signal restricted to the $N$ allocated subcarriers can be written as
\begin{equation}
	\underbrace{\begin{bmatrix}\mathbf{y}_1 \\ \mathbf{y}_2 \\ \vdots \\ \mathbf{y}_N \end{bmatrix}}_{\bar{\mathbf{y}}}
	\;=\;
		\underbrace{\operatorname{blkdiag}\begin{pmatrix}\mathbf{H}_{1} \\ \mathbf{H}_{2} \\ \vdots \\ \mathbf{H}_{ N}\end{pmatrix}}_{\bar{\mathbf{H}}}\,
	\bar{\mathbf{x}} \;+\; \bar{\mathbf{w}},
	\label{eq:received_signal}
\end{equation}
where $\mathbf{y}_n \in \mathbb{C}^{M \times 1}$ is the receive vector at subcarrier $n$, $\bar{\mathbf{H}} \in \mathbb{C}^{NM \times NK}$ is the block-diagonal MU channel across subcarriers, $\tilde{\mathbf{H}}=\bar{\mathbf{H}}(\mathbf{F}_N \otimes \mathbf{I}_K) \in \mathbb{C}^{NM \times NK}$ is the equivalent channel seen by $\bar{\mathbf{s}}$, and $\bar{\mathbf{w}} \sim \mathcal{CN}(\mathbf{0},\,\sigma^2 \mathbf{I}_{NM})$ is spatially and spectrally white additive Gaussian noise.The per-subcarrier MU channel matrix $\mathbf{H}_n \in \mathbb{C}^{M \times K}$ has entries $[\mathbf{H}_n]_{m,k}$ representing the complex baseband gain between UE $k$ and receive antenna $m$ on subcarrier $n$. 
	\vspace{-7pt}
\section{Power and Spectral Efficiency Gains of Non-Orthogonal Transmission with DFT-s-OFDM}
\label{sec:efficiency}

This section investigates two complementary benefits offered by DFT-s-OFDM-based non-orthogonal transmissions in the uplink system described in \S~\ref{sec:system_model}. First, we quantify the spectral-efficiency gains that arise when the number of concurrently transmitting UEs exceeds the number of base station antennas ($K > M$), showing that increasingly aggressive overloading (e.g., $K/M > 2$) yields correspondingly larger capacity gains. Second, we characterize the resulting power-efficiency improvements, which arise from three sources: the inherently low PAPR of DFT-s-OFDM relative to OFDM, the additional PAPR reduction obtained when each UE employs a low-order modulation, and a reduction in the number of receive antennas and RF chains required at the base station.
\vspace{-7pt}
\subsection{Capacity Gains}
\label{ssec:capacity}

Assuming equal per-UE transmit power $P_t$ and  perfect channel state information at the receiver, 
the information-theoretic sum capacity of the system in~\eqref{eq:received_signal} is given by
\begin{equation}
	C = \frac{1}{N}\sum_{n=1}^{N}
	\log_2\det\!\Bigl(\mathbf{I}_M
	+ \frac{P_t}{\sigma^2}\,\mathbf{H}_n\mathbf{H}_n^H\Bigr)
	\quad [\text{bits/s/Hz}],
	\label{eq:capG}
\end{equation}
where the factor $1/N$ averages over the allocated subcarriers. We note that~\eqref{eq:capG} remains valid in both the under-loaded ($K \leq M$) and overloaded ($K > M$) regimes, since the $M \times M$ Wishart-type product $\mathbf{H}_n\mathbf{H}_n^H$ is well-defined regardless of the relation between $K$ and $M$.

Two distinct gain mechanisms can be identified:
\begin{itemize}

 \item \emph{Spatial multiplexing gain.}
	When $K \leq M$, the system can resolve up to $\min(K,M) = K$ independent streams, and the sum rate scales linearly with $K$ at high SNR.
	
\item \emph{Received-power gain in the overloaded regime.}
	When $K > M$, additional independent spatial streams cannot be resolved; however, each additional UE still contributes its transmit power $P_t$ to the aggregate received signal. Since every column of $\mathbf{H}_n$ adds an independent rank-one contribution to $\mathbf{H}_n\mathbf{H}_n^H$, the eigenvalues of the Wishart matrix grow with $K$, increasing the argument of the log-determinant. This received-power gain is most pronounced in the low-SNR regime, where the capacity is approximately linear in SNR and therefore benefits directly from a higher aggregate received power.
\end{itemize}

\vspace{-10pt}
\subsection{Power Efficiency Gains}

The capacity expression in~\eqref{eq:capG} depends on the per-UE transmit power $P_t$, which in practice is constrained by the power amplifier (PA). A PA has a fixed saturation power $P_{\mathrm{sat}}$, and to avoid nonlinear distortion and out-of-band emissions, the mean transmit power must be backed off from saturation by at least the signal PAPR:
$P_{t_\text{dB}} = P_{{\mathrm{sat}}_\text{dB}} - \mathrm{BO}_{\mathrm{wf}},$	
where $\mathrm{BO}_{\mathrm{wf}}$ ($\mathrm{BO}_{\mathrm{wf}} \geq \mathrm{PAPR}$ in dB) denotes the waveform-dependent back-off. Because DFT-s-OFDM produces a near-single-carrier transmit signal, its PAPR is substantially lower than that of conventional OFDM for the same constellation and resource allocation (see Table~\ref{tab:PAPR}). Denoting the PAPR reduction relative to OFDM by
$	\Delta $,
the same PA can support a transmit power that is $\Delta$~dB higher under DFT-s-OFDM than under OFDM. The PAPR of DFT-s-OFDM is reduced even further with low-order modulations, and becomes especially small for QPSK and near-constant-envelope schemes such as $\pi/2$-BPSK (see Table~\ref{tab:PAPR}). This property motivates aggressive non-orthogonal transmissions with high overloading factors (e.g., $K/M>2$) that exceed what existing NOMA schemes can support.
Since each of the $K$ concurrently transmitting UEs independently benefits from the reduced back-off, the received-power gain at the base station is $\Delta$~dB, amplifying the capacity advantage identified in \S~\ref{ssec:capacity}. The additional transmit power also extends coverage. Consider a standard log-distance path loss model
$	\mathrm{PL}(d) = \alpha + 10\,n_p\log_{10}(d),$
where $\alpha$ is the intercept (including antenna gains and reference-distance loss) and $n_p$ is the path loss exponent. Let $d_{\max}$ denote the maximum range at which the received SNR meets the minimum required for reliable decoding. A $\Delta$~dB increase in $P_t$ translates directly into a $\Delta$~dB increase in the tolerable path loss, yielding a coverage range extension factor of
\begin{equation}
	\frac{d_{\max}^{(\mathrm{DFT\text{-}s})}}{d_{\max}^{(\mathrm{OFDM})}} = 10^{\,\Delta/(10\,n_p)}.
	\label{eq:range}
\end{equation}
For example when  $\Delta=3$~dB, and $n_p=2$ the coverage range gain is $41\%$ and area gain is $ 100\%$.

Beyond these per-UE gains, fully loaded ($K = M$) and overloaded ($K > M$) configurations approach the target capacity with significantly fewer RF chains and antennas than conventional massive MIMO systems, which rely on $K \ll M$ to compensate for the limitations of low-complexity linear receivers. The resulting reduction in active hardware translates into lower base-station power consumption and cost, complementing the UE-side efficiency gains discussed above.

%

	\vspace{-3pt}
\section{\fontsize{9.8}{9.8}\selectfont Nonlinear Receiver Processing  for  DFT-s-OFDM based Non-Orthogonal Transmissions}
\label{s:NLrec}

Realizing the capacity gains promised in  \S~\ref{ssec:capacity} hinges on the receiver design. To this end, we propose a receiver grounded in optimal receiver design principles that retains the same capacity-gain trends.

In systems that employ soft channel decoding, such as those specified in the 3GPP and Wi-Fi standards, maximum a posteriori (MAP) detection is optimal for minimizing the detection bit error probability \cite{STS}. The MAP detection process involves calculating log-likelihood ratios (LLRs). 
In the following, we introduce the soft information computation problem for receiver processing when DFT-s-OFDM based non-orthogonal transmissions are employed. 

The calculated soft information for each of the transmitted bits is expressed in terms of the LLR, which for the $i^{th}$ bit of user $l$ of the  $n^{th}$ subcarrier ($b_{i,l,n}$) is defined as \cite{STS} 
\begin{equation}
L(b_{i,l,n}) \triangleq  \ln{\bigg(\frac{P[b_{i,l,n}=1|\bar{\mathbf{y}},\tilde{\mathbf{H}}]}{P[b_{i,l,n}=0|\bar{\mathbf{y}},\tilde{\mathbf{H}}]}\bigg)}.
\label{eq:LD}
\end{equation}

\noindent After employing the max-log approximation, and equiprobable bits, 
 the LLR value of $b_{i,l,n}$ can be calculated as \cite{softAPPmimo,STSsoftin} 
\begin{align}
L(b_{i,l,n}) \approx & \min_{\bar{\mathbf{s}}\in S_{i,l,n}^{KN,0}}\big\{d(\bar{\mathbf{s}})\big\}-\min_{\bar{\mathbf{s}}\in S_{i,l,n}^{KN, 1}}\big\{d(\bar{\mathbf{s}})\big\},
\label{eq:LLR}
\end{align}
where 
\begin{align}
	d(\bar{\mathbf{s}}) =\frac{1}{\sigma^2}\|\bar{\mathbf{y}}-\tilde{\mathbf{H}}\bar{\mathbf{s}}\|^2,
	\label{eq:dsHSO}
\end{align}
and $S_{i,l,n}^{KN,0},S_{i,l,n}^{KN,1}$ are the subsets of symbol vectors having the $i^{th}$ bit of user $l$ taking the values $0,1$ respectively at the $n^{th}$ subcarrier.


In conventional OFDM systems, joint multi-user detection decouples across subcarriers and need only be performed over the $K$
users on each subcarrier. The minimization in~\eqref{eq:LLR}, by contrast, couples all $N$ subcarriers and $K$ users, leading to a complexity that grows exponentially in $KN$ and is therefore intractable in practice.

	\vspace{-6pt}
\section{Reduced Complexity Nonlinear Receiver Processing for DFT-s-OFDM}

Building on the receiver formulation of \S~\ref{s:NLrec}, we now develop a reduced-complexity framework for DFT-s-OFDM-based aggressive non-orthogonal transmissions that closely approximates its performance while remaining tractable in practice.

Because the DFT spreading operation couples symbols across subcarriers, conventional MU-MIMO and NOMA detection techniques cannot be applied directly to DFT-s-OFDM. Tree-search-based receivers, however, are known to closely approximate max-log MAP optimal processing in conventional OFDM-based MU-MIMO systems, which motivates their adaptation here. To adopt such receivers for DFT-s-OFDM, we split the receiver processing across two domains: interference cancellation and equalization are carried out in the frequency domain to reduce the equivalent channel to a single-tap flat-fading one, while demodulation and soft-information computation are performed in the time domain. By decoupling processing across subcarriers in frequency and across multipath taps in time, this approach avoids the complexity of joint detection in either domain. The underlying principles apply to any tree-search-based detector or successive interference cancellation receiver.

First, to transform the minimization problems in~\eqref{eq:LLR} to an equivalent tree search, and to reduce complexity, we perform a QR decomposition on each component matrix $\mathbf{H}_{\scriptsize n}$ of  the block diagonal $\bar{\mathbf{H}}$ matrix in~\eqref{eq:received_signal}.
\begin{equation}
 \check{\mathbf{H}}_{ n}\triangleq\begin{bmatrix} \mathbf{H}_{ n} \\ \lambda\mathbf{I}_{K} \end{bmatrix} = \check{\mathbf{Q}}_{n}\mathbf{R}_{n}=\begin{bmatrix} \mathbf{Q}_{1,n} \\  \mathbf{Q}_{2, n} \end{bmatrix} \mathbf{R}_{ n}, 
\end{equation}
 
\noindent with the regularization parameter $\lambda=\sigma/\sqrt{P_t}$, $ \mathbf{Q}_{1,n}\in \mathbb{C}^{M \times K}$  is the first $M$ rows of the unitary matrix $\check{\mathbf{Q}}_{n}$ and $\mathbf{R}_{n}\in \mathbb{C}^{K \times K}$ is an upper triangular matrix. This regularization makes the QR decomposition and the following tree search applicable to scenarios where $K>M$. Then,~\eqref{eq:received_signal} can be expressed as

\begin{equation}
\begin{bmatrix}\mathbf{{y}}_{1} \\ \mathbf{{y}}_{ 2}  \\ \vdots \\ \mathbf{{y}}_{N}\end{bmatrix}=\operatorname{blkdiag}\begin{pmatrix}\mathbf{Q}_{1, 1}\mathbf{R}_{ 1} \\ \mathbf{Q}_{1,2}\mathbf{R}_{2} \\ \vdots \\ \mathbf{Q}_{1, N}\mathbf{R}_{ N}\end{pmatrix}\bar{\mathbf{x}}+\bar{\mathbf{w}}.
\end{equation}
Then, by performing the matrix vector multiplication $\mathbf{Q}_{1, n}^H\mathbf{y}_{ n}$ to obtain  $\mathbf{\tilde{y}}_{n}$ for each $\mathbf{y}_{ n}$, the equivalent received observables are
\begin{equation}
\begin{bmatrix}\mathbf{\tilde{y}}_{1} \\ \mathbf{\tilde{y}}_{ 2}  \\ \vdots \\ \mathbf{\tilde{y}}_{N}\end{bmatrix}=\operatorname{blkdiag}\begin{pmatrix}\mathbf{R}_{ 1} \\ \mathbf{R}_{2} \\ \vdots \\ \mathbf{R}_{ N}\end{pmatrix}\bar{\mathbf{x}}+\tilde{\mathbf{w}}.
\end{equation}
\noindent Note that due to $ \check{\mathbf{Q}}_{ n}$ being a unitary matrix, the variance of $\tilde{\mathbf{w}}$ remains the same as that of $\bar{\mathbf{w}}$.

The approach comprises four main stages: channel equalization (\S \ref{ss:ChannelEq}), demodulation (\S \ref{ss:Demod}), and interference cancellation (\S \ref{ss:IC}) are applied successively to all $K$ users, starting from user $K$, followed by a final stage in which soft information is computed from the distance metrics (\S \ref{ss:LLRcomp}) recursively accumulated during the demodulation stage.
\vspace{-10pt}
\subsection{Channel Equalization (Frequency-Domain):}
\label{ss:ChannelEq}
Since, the $\mathbf{R}_n$ matrices are upper triangular, starting from user $K$ channel equalization can be conducted successively in the frequency domain. In general, for the $k^{th}$ user the channel equalized received observable $\mathbf{\tilde{x}}_{k}$ is
\begin{equation}
\begin{bmatrix} \tilde{x}_{k_1} & \tilde{x}_{k_2} & \hdots & \tilde{x}_{k_N} \end{bmatrix}=\begin{bmatrix} \frac{\hat{y}_{k_1}}{R_{k,k,1}} & \frac{\hat{y}_{k_2}}{R_{k,k,2}} & \hdots & \frac{\hat{y}_{k_N}}{R_{k,k,N}}\end{bmatrix}
\end{equation}
where 
\begin{equation}
\begin{bmatrix} \hat{y}_{K_1} & \hat{y}_{K_2} & \hdots & \hat{y}_{K_N}\end{bmatrix} = \begin{bmatrix} \tilde{y}_{K_1} & \tilde{y}_{K_2} & \hdots & \tilde{y}_{K_N}\end{bmatrix} 
\end{equation}
\noindent is the equivalent received signal following interference cancellation (\S \ref{ss:IC}).
\vspace{-10pt}
\subsection{Demodulation (Time-Domain):}
\label{ss:Demod}
Following channel equalization demodulation can be performed in the time domain based on
\begin{equation}
\mathbf{\tilde{s}}_{k}=\mathbf{F}^H_N  \mathbf{\tilde{x}}_{k}
\end{equation}
Then, $\mathbf{\tilde{s}}_{k}$ can be demodulated to constellation symbols in $\mathcal{O}$ as  $\begingroup 
\setlength\arraycolsep{1pt}\begin{bmatrix} \hat{s}_{k_1} & \hat{s}_{k_2} & \dots & \hat{s}_{k_N} \end{bmatrix}=\begin{bmatrix} \lfloor \tilde{s}_{k_1} \rceil & \lfloor \tilde{s}_{k_2} \rceil & \dots & \lfloor \tilde{s}_{k_N} \rceil \end{bmatrix}\endgroup$, where $\lfloor \tilde{s}_{k_n} \rceil$ denotes rounding $\tilde{s}_{k_n}$ to the nearest constellation symbol in $\mathcal{O}$ when successive interference cancellation receiver processing is employed. Additionally, when a tree search based detection scheme such as Fixed Complexity Sphere Decoder (FCSD) \cite{newsoftfsd,SFSD} or MultiSphere massively parallel nonlinear (MPNL) processing \cite{nikitopoulos2024towards,nikitopoulos2018massively} is employed, demapping $\tilde{s}_{k_n}$ to $\hat{s}_{k_n}$ depends on the particular treepath and level (user) $k$. For example, FCSD uses a predefined tree path to symbol demapping as discussed in \cite{SFSD}. In contrast, MPNL processing identifies the most promising tree paths based on upper triangular matrix $\mathbf{R}_n$. Specifically, a tree path is represented by a relative distance vector $\mathbf{b}_n$ with integer elements  $b_{k_n} \in [1,|\mathcal{O}|], k \in [1,K]$ that indicate the  order of the distance relative to  $\tilde{s}_{k_n}$. Then a probabilistic metric can be defined based solely on $\mathbf{R}_n$ as 
\begin{align}
	\mathcal{M}\left(\mathbf{b}_n\right)=\sum_{k=1}^{K} \alpha_k \left[b_{k_n}-1\right] \left| R_{k,k,n}\right|^2  \thickspace \thickspace  \forall n \in[1 \thinspace N]
\end{align}
where $\alpha_k$ depends on the minimum distance between constellation symbols. For example, $\alpha_k=1.11$ when  the minimum distance is $2$. Then, identifying the most promising tree paths correspond to the $\mathbf{b}_n$ that minimize $\mathcal{M}\left(\mathbf{b}_n\right)$.
This preprocessing stage does not depend on the received signal \(\mathbf{y}_{n}\) and need only be performed when the channel estimates are updated.
Then, these tree paths are used for demapping $\tilde{s}_{k_n}$ to $\hat{s}_{k_n}$ based on the relative distance to $\tilde{s}_{k_n}$. Hence, $\hat{s}_{k_n}$ is not restricted to the closest symbol to $\tilde{s}_{k_n}$ it may be the $2^{nd}$, $3^{rd}$ or up to the  $|\mathcal{O}|^{th}$ closest.
The associated distance metric can be computed recursively for each tree path as

\begin{equation}
	\begin{aligned}
		d(\hat{\mathbf{s}}_{k_n}) &= d(\hat{\mathbf{s}}_{{k+1}_n}) 
		+ \frac{\left| R_{k,k,n}\right|^2}{\sigma^2}\,\bigl|\tilde{s}_{k_n} - \hat{s}_{k_n}\bigr|^2 
		- \frac{\lambda^2}{\sigma^2}\,\bigl|\hat{s}_{k_n}\bigr|^2, \\
		&\quad \forall\, n \in [1 \thinspace N],\ k < K, 
		\quad \text{with } d(\hat{\mathbf{s}}_{{K+1}_n}) = 0.
	\end{aligned}
\end{equation}

\noindent To simplify notation  $d(\hat{\mathbf{s}}_{1_n})$ will be denoted as  $d(\hat{\mathbf{s}}_{n})$. Then, to perform interference cancellation in the frequency domain, DFT spreading is applied to the demodulated symbols as
\begin{equation}
\mathbf{\hat{x}}_{k}=\mathbf{F}_N  \mathbf{\hat{s}}_{k}
\end{equation}
\subsection{Interference Cancellation (Frequency-Domain):}
\label{ss:IC}
Based on $\mathbf{\hat{x}}_{k}$, computed in the demodulation step, interference cancellation can be performed on the equivalent received observable of user $k$ as
\begin{equation}
\hat{y}_{k_n}=\tilde{y}_{k_n}-\sum_{l=k+1}^K\hat{x}_{l_n}R_{k,l,n}  \quad\thickspace \thickspace \forall n \in[1 \thinspace N],\thinspace k \in[1 \thinspace K]
\end{equation}
\noindent For, tree search based processing, the above steps need to be performed for each processed tree path. For example, MPNL typically processes 16 tree paths. 

\subsection{Soft Information Computation:}
\label{ss:LLRcomp}
Finally, a list of most promising symbol vectors (i.e., $\mathbf{\hat{s}}_{n}=\mathbf{\hat{s}}_{1_n}$) and their corresponding distance metrics (i.e., $d(\mathbf{\hat{s}}_{n})$) are generated for all $n \in[1 \thinspace N]$, which can be used to compute LLRs as discussed in~\eqref{eq:LLR}  as
\begin{align}
	L(b_{i,l,n}) \approx & \min_{\mathbf{\hat{s}}_{n}\in S_{i,l}^{K,0}}\big\{d(\mathbf{\hat{s}}_{n})\big\}-\min_{\mathbf{\hat{s}}_{n}\in S_{i,l}^{K, 1}}\big\{d(\mathbf{\hat{s}}_{n})\big\},
	\label{eq:LLRred}
\end{align}
where $S_{i,l}^{K,0},S_{i,l}^{K,1}$ are the subsets of symbol vectors having the $i^{th}$ bit of user $l$ taking the values $0,1$ respectively. In contrast to the LLR computation in~\eqref{eq:LLR} where the minimization is across subcarriers,~\eqref{eq:LLRred} enables independent processing for each subcarrier, thus substantially reducing complexity while preserving optimality. 
The equivalence of~\eqref{eq:LLR} and~\eqref{eq:LLRred} follows in three steps, each invoking the unitary invariance of the Euclidean norm. First, since $\check{\mathbf{Q}}_n$ is unitary, $\|\mathbf{y}_n-\mathbf{H}_n\mathbf{x}_n\|^2=\|\tilde{\mathbf{y}}_n-\mathbf{R}_n\mathbf{x}_n\|^2-\lambda^2\|\mathbf{x}_n\|^2+\|\mathbf{Q}_{c,n}^H\mathbf{y}_n\|^2$, where $\mathbf{Q}_{c,n}$ denotes the complementary block of the full QR decomposition; the last term is symbol-independent and absorbed into a constant $c$. Second, applying the same argument to $\mathbf{F}_N$ and summing over the $N$ subcarriers yields $\sum_{n=1}^{N}\tfrac{1}{\sigma^2}\|\mathbf{y}_n-\mathbf{H}_n\mathbf{x}_n\|^2=\sum_{n=1}^{N}d(\hat{\mathbf{s}}_n)+c$. 
Finally, since $\sum_{n=1}^{N}\tfrac{1}{\sigma^2}\|\mathbf{y}_n-\mathbf{H}_n\mathbf{x}_n\|^2=\tfrac{1}{\sigma^2}\|\bar{\mathbf{y}}-\tilde{\mathbf{H}}\bar{\mathbf{s}}\|^2$, we obtain $d(\bar{\mathbf{s}})=\sum_{n=1}^{N}d(\hat{\mathbf{s}}_n)+c$.

\vspace{-6pt}
\section{Evaluations}
\label{s:Eval}
In this section, we evaluate the proposed scheme (DFT-s-NL) against existing baselines in terms  of achievable spectral efficiency (SE), computational complexity, and power efficiency \footnote{The simulation setup follows the 3GPP TDL-A profile with 48 active subcarriers at 30 kHz spacing, a 100 ns delay spread, and a 2 GHz carrier}.  To ensure a fair comparison, we extend the conventional LMMSE detector to DFT-s-OFDM-based MIMO following the same design principles as DFT-s-NL: a per-user IDFT stage after frequency-domain matrix inversion, with demodulation and soft-information computation performed in the time domain. Since no detector for overloaded ($K > M$) DFT-s-OFDM transmissions has been reported in the literature to the best of our knowledge, we further construct a successive interference cancellation baseline (DFT-s-SIC) built on the same principles. The evaluation proceeds in three steps. First, we demonstrate the SE and power-efficiency gains of aggressive non-orthogonal transmission by supporting more UEs than receive antennas. Second, we characterise the complexity versus performance trade-off of DFT-s-NL against state-of-the-art detectors for DFT-s-OFDM-based MU-MIMO systems. Finally, we assess robustness under practical operating conditions, including realistic channel estimation with 3GPP reference signals and high-mobility scenarios representative of the Doppler spread encountered on LEO satellite platforms.

\begin{figure}[!ht]
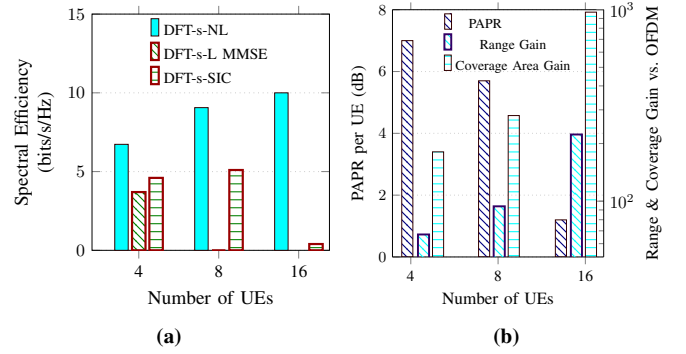

	\begin{subfigure}{.49\columnwidth}
		\centering
		\includegraphics[width=0.98\columnwidth]{figures/SEaggressive2.tikz}	
		\caption{}
		\label{fig:SEA}
	\end{subfigure}
	\begin{subfigure}{.49\columnwidth}
		\centering
		\includegraphics[width=0.98\columnwidth]{figures/PAPR.tikz}	
		\caption{}
		\label{fig:SEUEs}
	\end{subfigure}
	\caption{\textbf{(a)} Spectral efficiency and \textbf{(b)} PAPR, along with the corresponding range and coverage gain, as the number of UEs increases beyond the number of antennas for a four-antenna receiver.}
	\label{fig:ratewithUEs}
		\vspace{-8pt}
\end{figure}

Fig. \ref{fig:SEA} plots the SE as the number of concurrent UEs grows beyond the number of antennas at 4 dB SNR. The Modulation and Coding Scheme (MCS) is chosen from the 3GPP tables to maximize throughput. We consider three scenarios: (i) a conventional $4\times4$ MU-MIMO baseline with DFT-s-OFDM, (ii) an overloaded case with eight UEs, and (iii) a more aggressively overloaded case with sixteen UEs. Consistent with the analysis in \S~III-A, each step into the overloaded regime delivers additional SE gains. Fig. \ref{fig:SEUEs} shows the corresponding PAPR and the range and coverage gains it implies. Because overloading allows each UE to fall back to a lower-order modulation, QPSK in scenario (ii) versus 16-QAM in (i), and $\pi/2$-BPSK in (iii),  the PAPR of the DFT-s-OFDM waveform decreases monotonically across scenarios. As discussed in \S~III-B, this PAPR headroom translates directly into range and coverage gains.

A second benefit emerges when comparing against an LMMSE-based MIMO system delivering equivalent throughput: scenarios (ii) and (iii) achieve the same performance with substantially fewer RF chains and antennas. The eight-UE case halves the antenna and RF-chain count, and proportionally the front-end power consumption at the base station, while the sixteen-UE case reduces them by a factor of four.

\begin{figure}[!ht]
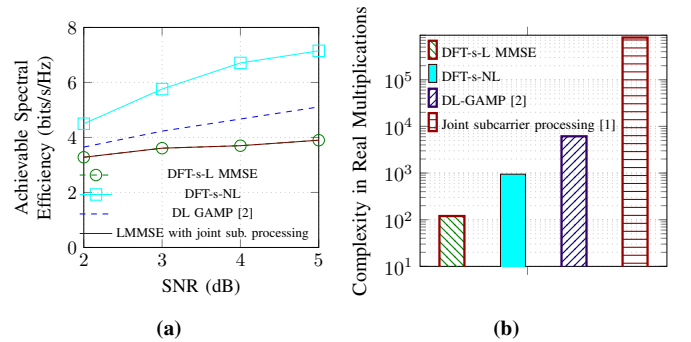

	\begin{subfigure}{.49\columnwidth}
		\centering
		\includegraphics[width=0.98\columnwidth]{figures/tp3.tikz}	
		\caption{}
	\label{fig:4x4tp}
	\end{subfigure}
	\begin{subfigure}{.49\columnwidth}
		\centering
		\includegraphics[width=0.98\columnwidth]{figures/comp2.tikz}	
		\caption{}
	\label{fig:4x4comp}
	\end{subfigure}
	\caption{\textbf{(a)} Achievable spectral efficiency, and the \textbf{(b)} corresponding complexity at 4 dB SNR, for a four-antenna receiver supporting four users. Perfect channel knowledge is assumed.}
		\vspace{-12pt}
\end{figure}

%

Fig. \ref{fig:4x4tp} compares the achievable SE of \systemname against existing DFT-s-OFDM detectors in a $4\times4$ MU-MIMO system. SE achievable by  \systemname exceeds that of DFT-s-L MMSE by nearly $2\times$ and the deep-learning Gaussian message-passing algorithm (DL-GAMP) \cite{DL_GAMP} by $40\%$. Fig. \ref{fig:4x4comp} shows the corresponding complexity in multiplications. The complexity of \systemname is only $15\%$ that of DL-GAMP, even before accounting for the latter's training overhead. On non-sparse channel matrices (e.g., $M=K$), DL-GAMP requires many iterations to reach acceptable error performance, and each iteration adds a per-user DFT/IDFT pair. Compared to DFT-s-L MMSE, \systemname incurs less than $8\times$ the complexity. The results also confirm that joint subcarrier processing is impractical even at the modest number of active subcarriers considered here. Table \ref{tab:Complexity_tab} summarises the complexity in multiplications for  DL-GAMP, DFT-s-L MMSE and \systemname.

\begin{table}[!ht]
	\centering
	\caption{Detection complexity, in multiplications, for an $M \times K$
		 DFT-s-OFDM system with $N$
		subcarriers, normalized per subcarrier and excluding DL-GAMP training. $T$
		denotes the number of GAMP iterations and $n_{tp}$ the number of MPNL tree paths.}
	\label{tab:Complexity_tab}
	\resizebox{\columnwidth}{!}{%
		\begin{tabular}{|l|c|c|}
			\hline
			\textbf{Detection method}
			& \textbf{Per received vector}
			& \makecell{\textbf{Per channel}\\\textbf{update}} \\
			\hline
			\makecell[l]{DL-GAMP}
			& $TK(4K+6\sqrt{|\mathcal{O}|}+4\log_2 N)$
			& - \\
			\hline
			\makecell[l]{DFT-s-L MMSE}
			& $3MK+2K\log_2 N$
			& $6M K^2$ \\
			\hline
			\makecell[l]{DFT-s-NL}
			& $3MK+n_{tp}K(2\log_2 N+3(K+(1+K)/2))$
			& $6M K^2$ \\
			\hline
		\end{tabular}%
	}
\end{table}

\begin{figure}[!ht]
	\centering
	\includegraphics[width=0.98\columnwidth]{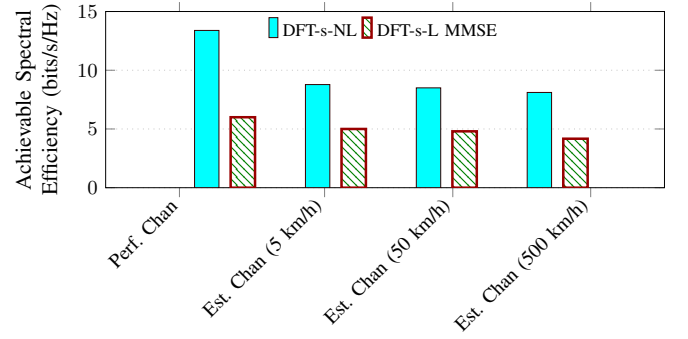}
	\caption{Eight-antenna receiver supporting eight users. Results assume a TDL-A channel with practical channel estimation using 3GPP reference signals.}
	\label{fig:8x8chanest}
		\vspace{-15pt}
\end{figure}

Fig. \ref{fig:8x8chanest} evaluates an $8\times8$ configuration, where eight UEs are served by an eight-antenna receiver, under practical channel estimation. 3GPP reference signals and frame structures are used to capture both estimation errors and the Doppler spread induced by high relative mobility \footnote{We assume UEs are synchronized to a single satellite using the 3GPP NTN time–frequency procedures, which rely on GNSS signaling, satellite ephemeris, and 3GPP reference signals \cite{3gpp.38.211}. Each UE obtains the ephemeris from the satellite's broadcast synchronization signal block and combines it with its GNSS-derived geolocation to pre-compensate for Doppler shift and propagation delay.}. We consider radial speeds from 5 km/h to 500 km/h, a range representative of LEO/MEO satellites operating near zenith, where radial velocity is minimal. The 500 km/h ceiling reflects the maximum mobility supported by current 3GPP reference-signal designs: the Demodulation Reference Signal (DM-RS) \cite{3gpp.38.211}, which carries the channel-estimation overhead, may occupy up to four OFDM symbols at high mobility. SNR is at 4 dB, and the MCS is chosen from the 3GPP tables to maximize throughput. Under these realistic conditions, \systemname delivers up to $2.2\times$ the SE of DFT-s-L MMSE while remaining robust to estimation errors, Doppler-induced impairments, and reference-signal overhead.

	\vspace{-5pt}
\section{Conclusions}

This paper introduced a framework that unifies the spectral efficiency gains of aggressive non-orthogonal transmission with the favorable PAPR characteristics of DFT-s-OFDM. The proposed scheme achieves up to twice the spectral efficiency of conventional systems while remaining robust to practical channel-estimation errors and the Doppler impairments encountered in high-mobility scenarios. Against recent nonlinear MIMO detectors, the scheme delivers 40\% higher spectral efficiency at 15\% of the complexity, establishing a favorable performance–complexity trade-off critical for deployment. It further reduces UE-side PAPR by up to 6 dB over DFT-s-OFDM-based MIMO and 11 dB over OFDM-based systems, directly relaxing power amplifier back-off requirements.

\bibliographystyle{IEEEtran}
	\vspace{-8pt}
\bibliography{reference}

@techreport{3gpp.38.211,
 author = {3GPP},
 institution = {{3rd Generation Partnership Project (3GPP)}},
 month = {Dec.},
 note = {Version 17.1.0},
 number = {38.211},
 title = {{Physical Channels and Modulation}},
 type = {Technical Specification (TS)},
 year = {2022}
}

@article{nikitopoulos2018massively,
  title={Massively Parallel Tree Search for High-Dimensional Sphere Decoders},
  author={Nikitopoulos, Konstantinos and Georgis, Georgios and Jayawardena, Chathura and Chatzipanagiotis, Daniil and Tafazolli, Rahim},
  journal={IEEE Trans. Parallel Distrib. Syst.},
  year={2018},
  publisher={IEEE}
}

@ARTICLE{softAPPmimo, 
 author={K. Nikitopoulos and G. Ascheid}, 
 journal={IEEE Trans. Veh. Technol}, 
 title={Approximate {MIMO} Iterative Processing With Adjustable Complexity Requirements}, 
 year={2012}, 
 volume={61}, 
 number={2}, 
 pages={639-650}, 
 month={Feb},}

@ARTICLE{STSsoftin, 
 author={Studer, C. and Bolcskei, H.}, 
 journal={IEEE Trans. Inf. Theory}, 
 title={Soft Input Soft Output Single Tree-Search Sphere Decoding}, 
 year={2010}, 
 volume={56}, 
 number={10}, 
 pages={4827-4842}, 
 doi={10.1109/TIT.2010.2059730}, 
 ISSN={0018-9448}, 
 month={Oct},}

@ARTICLE{STS, 
 author={Studer, C. and Burg, A. and Bolcskei, H.}, 
 journal={IEEE J. Sel. Areas Commun.}, 
 title={Soft-output sphere decoding: algorithms and {VLSI} implementation}, 
 year={2008}, 
 volume={26}, 
 number={2}, 
 pages={290-300}, 
 ISSN={0733-8716}, 
 month={Feb},}

@INPROCEEDINGS{SFSD,
  author={Barbero, Luis G. and Ratnarajah, T. and Cowan, Colin},
  booktitle={Proc. of IEEE ICASSP}, 
  title={A Low-Complexity Soft-{MIMO} Detector Based on the Fixed-Complexity Sphere Decoder}, 
  year={2008},
  volume={},
  number={},
  pages={2669-2672},
  doi={10.1109/ICASSP.2008.4518198}
	}

@INPROCEEDINGS{newsoftfsd,

  author={Dai, Ya-Xin and Jhang, Shih-Jie and Chen, Yen-Ming and Lan, Sheng-Ping and Ueng, Yeong-Luh},

  booktitle={Proc. of IEEE ACSSC}, 

  title={An Efficient Soft Output {MIMO} Detector Architecture Considering High-order Modulations}, 

  year={2022},

  volume={},

  number={},

  pages={623-627},

  doi={10.1109/IEEECONF56349.2022.10051955}}

@ARTICLE{SCMA,
  author={Yu, Lisu and Liu, Zilong and Wen, Miaowen and Cai, Donghong and Dang, Shuping and Wang, Yuhao and Xiao, Pei},
  journal={IEEE Commun. Stand. Mag.}, 
  title={{Sparse Code Multiple Access for 6G Wireless Communication Networks: Recent Advances and Future Directions}}, 
  year={2021},
  volume={5},
  number={2},
  pages={92-99},
  doi={10.1109/MCOMSTD.001.2000049}}

@ARTICLE{SCMA2,
  author={Rebhi, Manel and Hassan, Kais and Raoof, Kosai and Chargé, Pascal},
  journal={IEEE Open Journal of the Communications Society}, 
  title={{Sparse Code Multiple Access: Potentials and Challenges}}, 
  year={2021},
  volume={2},
  number={},
  pages={1205-1238},
  doi={10.1109/OJCOMS.2021.3081166}}

@article{nikitopoulos2024towards,
  author       = {K. Nikitopoulos and G. N. Katsaros and M. Filo and C. Jayawardena and R. Tafazolli},
  title        = {{Towards Software-Based, MIMO, Open-RAN PHY Architectures with both Linear and Non-Linear Processing}},
  journal      = {IEEE Commun. Mag.},
  year         = {2024}
}

@ARTICLE{DL_GAMP,
  author={Zeng, Yuwei and Ge, Yingmeng and Tan, Xiaosi and Ji, Zhenhao and Zhang, Zaichen and You, Xiaohu and Zhang, Chuan},
  journal={IEEE Trans. Veh. Technol.}, 
  title={{A Deep-Learning-Aided Message Passing Detector for MIMO SC-FDMA}}, 
  year={2024},
  volume={73},
  number={7},
  pages={10767-10771},
  doi={10.1109/TVT.2024.3366244}}

@ARTICLE{SC-FDMblccirc,
  author={Jeong, Dahoon and Kim, Jaekwon},
  journal={IEEE Trans. Veh. Technol.}, 
  title={{Signal Detection for MIMO SC-FDMA Systems Exploiting Block Circulant Channel Structure}}, 
  year={2016},
  volume={65},
  number={9},
  pages={7774-7779},
  doi={10.1109/TVT.2015.2485264}}

\end{document}